# Origin of multistate resistive switching in Ti/manganite/SiO$_x$/Si heterostructures

W. Román Acevedo[1,2], C. Acha[2,3], M.J. Sánchez[2,4], P. Levy[1,2] and D. Rubi[1,2,5,*]

1. Gerencia de Investigación y Aplicaciones, CNEA, Av. Gral Paz 1499 (1650), San Martín, Buenos Aires, Argentina
2. Consejo Nacional de Investigaciones Científicas y Técnicas (CONICET), Godoy Cruz 2290 (1425), Buenos Aires, Argentina.
3. Depto. de Física, FCEyN, Universidad de Buenos Aires & IFIBA-CONICET, Pab I, Ciudad Universitaria, Buenos Aires (1428), Argentina
4. Centro Atómico Bariloche and Instituto Balseiro, 8400 San Carlos de Bariloche, Río Negro, Argentina
5. Escuela de Ciencia y Tecnología, UNSAM, Campus Miguelete (1650), San Martín, Buenos Aires, Argentina

We report on the growth and characterization of Ti/La$_{1/3}$Ca$_{3/2}$MnO$_3$/SiO$_2$/n-Si memristive devices. We demonstrate that using current as electrical stimulus unveils an intermediate resistance state, in addition to the usual high and low resistance states that are observed in standard voltage controlled experiments. Based on thorough electrical characterization (impedance spectroscopy, current-voltage curves analysis), we disclose the contribution of three different microscopic regions of the device to the transport properties: an ohmic incomplete metallic filament, a thin manganite layer below the filament tip exhibiting Poole-Frenkel like conduction, and the SiO$_x$ layer with an electrical response well characterized by a Child-Langmuir law. Our results suggest that the existence of the SiO$_x$ layer plays a key role in the stabilization of the intermediate resistance level, indicating that the combination of two or more active resistive switching oxides adds functionalities in relation to single-oxide devices. We understand that these multilevel devices are interesting and promising as their fabrication procedure is rather simple and they are fully compatible with standard Si-based electronics.

* Corresponding author (DR), diego.rubi@gmail.com

The continuous increase of memory densities over the last decades has been, to a large extent, responsible for the impressive evolution of new micro- and nanoelectronics devices. Further progress, especially in mobile appliances, crucially depends on new developments in solid-state memories. New memory devices should combine non-volatility, speed, durability and extended scaling. Different technologies have been proposed to accomplish these requirements, like phase-change memories, magnetic or ferroelectric random access memories and resistive random access memories (RRAM). The latter are based on metal/insulator/metal or metal/insulator/semiconductor structures, usually called "memristors", that display a significant, non-volatile, change in their resistance upon application of electrical stress. This effect was named resistive switching (RS) from early 00's (see reviews [1,2] and references therein). RS has been found to ubiquitously exist in a huge variety of simple and complex transition metal oxides, and shows promising properties in terms of scalability, low power consumption and fast write/read access times. In addition, memristors were shown to behave similarly to human brain bit-cells (synapses) [3, 4], suggesting the possibility of developing disruptive devices with neuromorphic behaviour. Thus, additional functionalities foster the search of selected oxides that could potentiate memory devices capabilities.

Massive Si planar technology prompts the use of heavily doped Si substrates as the bottom electrode (instead of a metal) in memristive stacks. Naturally, $SiO_x$ was tested as an active RS oxide with promising results. Metal/$SiO_x$/Si structures have been shown to display resistive switching even for ultrathin $SiO_x$ layers (~1nm thick) [5]. The associated physical mechanisms were related to the formation of conducting nanofilaments, either due to the reduction of Si atoms [6,7], to the drift/ diffusion of the metallic electrode through the oxide layer [8]. These devices displayed excellent retention time, high ON/OFF ratio and multilevel storage potential [9,10]. Other reports account for a localized switching model involving proton exchange reactions, where resistive switching is related to an interplay between conductive Si-H-Si and non-conductive H-SiSi-H defects. [11-14]. This effect was found to be stable at high temperatures [11,14]. On the other hand, it has been proposed that inserting a $SiO_x$ layer in contact with other active memristive oxide such as ZnO, changes the resistive switching behavior from bipolar to complementary (similar to the behavior of two serial RRAM devices with bipolar resistive switching in a back-to-back structure) [15].

This effect seems to be strongly related to the oxygen storage capacity of the inserted $SiO_x$ layer. A related scenario was proposed for $TiN/SiO_2/Fe$ structures, where the presence of $SiO_2$ allows the spontaneous formation of $FeO_x$ during the Fe deposition, which is ultimately responsible of the memristive behavior [16]. These works suggest that the RS behavior can be dramatically affected by combining two or more active oxides.

On the other hand, memristive behaviour in manganese oxides, usually known as "manganites", has been extensively studied in the last 15 years since the initial report of Ignatiev et al. [17]. Phenomenological models that successfully describe the electrical behavior have been developed [18,19]. Proposed mechanisms associated to the memristive behavior in manganites include the modulation of the height of the Schottky barrier at the oxide/metal interface [20], the oxidation/reduction of a thin layer of the metallic electrode in contact with the manganite [22] and the creation/disruption of conducting (metallic) filaments bridging both electrodes [20,22]. Concerning electronic transport, different mechanisms such as Poole-Frenkel (PF) [23], space charged limited current (SCLC) [24] or Schottky (Sch) conduction [25,26], have been reported for manganite-based devices.

We have already faced the study of memristive $TE/La_{2/3}Ca_{1/3}MnO_3$ (TE: Ti/Cu with thicknesses of 10 and 100nm, respectively) devices grown on heavily doped n-Si (which acted as bottom electrode) without removing the native $SiO_x$ layer [20]. It was found a bipolar RS behavior with a crossover between two of the above mentioned mechanisms -modulation of the metal/manganite interface resistance and metallic filament formation- controlled by the compliance current (CC) programmed during the transition from high to low resistance states (SET process). We recall that in that work the applied electrical stress was *voltage* and only two stable resistance states ($R_{HIGH}$ and $R_{LOW}$) were obtained. We argued that the time window of a few ms that standard source-meters take to stabilize the CC, leads to an uncontrolled power overshoot during the SET process (power $P=V^2/R$) that induces the hard dielectric breakdown of the $SiO_x$ layer laying between the Si substrate and the manganite layer [27,20]. This unwanted effect could be avoided if the electrical stress is in current controlled mode because the dissipated power during the SET process remains self limited ($P=I^2R$). This strategy has been followed in Ref. [28], where an intermediate resistance state ($R_{INT}$) was unveiled. This multilevel behavior could be related to a combination of the oxidation/reduction of the $SiO_x$ ultrathin layer plus the formation of a (partial) metallic filament which does not bridge both electrodes [28,29]. However, further evidence clarifying this physical mechanism is still needed.

In the present work, we report on impedance spectroscopy (IS) experiments performed on the different resistive states of $Ti/La_{1/3}Ca_{2/3}MnO_3/SiO_x/n$-Si structures. Complemented with a careful analysis of the dynamical current-voltage curves, the IS analysis allows us to conclusively setting the physical scenario associated with the multilevel resistive switching observed in these systems, in which the capability of the $SiO_x$ layer for taking and releasing oxygen ions from and to the nearby manganite layer is proposed to play a key role. These results confirm that the combination of two active oxides improves the functionalities of the devices with respect to single-oxide structures.

$La_{1/3}Ca_{2/3}MnO_3$ manganite (LCMO) thin films were grown on top of heavily doped n-type silicon ($\delta<5m\Omega cm$) by pulsed laser deposition. No chemical removal of the native $SiO_x$ layer was performed. The thickness of this layer in an as-received substrate was estimated in ~1nm by X-ray photoemission spectroscopy (see Supplementary Information), although it may become thicker after heating the substrate in $O_2$ atmosphere in the chamber, prior to deposition. A 266nm Nd:YAG solid state laser, operating at a repetition frequency of 10Hz, was used. The deposition temperature and oxygen pressure were 850ºC and 0.13mbar, respectively. The films resulted single phase and polycrystalline [28]. The film thickness was estimated by cross-section scanning electron microscopy imaging in 100nm. Ti top electrodes (100nm thick) were deposited by sputtering and shaped by means of optical lithography. Top electrode areas ranged between $32\times10^3$ μm$^2$ and $196\times10^3$ μm$^2$. Electrical characterization was performed at room temperature with a Keithley 2612 source-meter hooked to a probe station. The n-type silicon substrate was grounded and used as bottom electrode. The electrical stimulus was applied to the top electrode. Samples were electrically formed by applying a positive current of 2mA, which starts the formation of a metallic filament and renders the device in a high resistance state [28]. Upon forming, the resistance changes from the virgin state $R_V$~2MΩ to $R_{HIGH}$~60kΩ. Complex impedance was measured by using a AutoLab PGSTAT302N impedance analyzer at room temperature. A fixed AC signal of 100 mV was applied, with frequencies between 10Hz and 1MHz.

Figures 1(a) and (b) display a dynamical current-voltage (I-V) curve and a hysteresis switching loop (HSL), respectively. The I-V curve is obtained by applying a sequence of current pulses of different amplitudes (0→17.5mA→-16mA→0, with a time-width of a few milliseconds and a step of 0.1mA) while the voltage is measured during the application of the pulse. Additionally, after each current pulse we apply a small undisturbing reading voltage of 100mV that allows measuring the current and evaluating the remnant resistance state (HSL). The device is initially in $R_{HIGH}$ and, upon positive stimulus, goes through two transitions, first to $R_{LOW}$ (SET process) after a pulse of 2mA and then to $R_{INT}$ after a pulse of 14mA. When the polarity of the stimulus is reversed, a transition from $R_{INT}$ to $R_{HIGH}$ (RESET process) is observed after a -16mA pulse. Also, the transition from $R_{LOW}$ to $R_{HIGH}$ was obtained for negative bias. The three (remanent) resistance states (60kΩ, 3.5kΩ and 800Ω) are clearly seen in the HSL of Figure 1(b). These resistance levels were stable for at least $10^4$ sec and the device was found to display a reproducible behavior for at least 70 cycles [28].

In order to gain information regarding the physical scenario behind the described phenomenology we have performed complex impedance spectroscopy on the three resistance levels. This technique allows to characterize as a function of the frequency the dielectric contribution from different regions of the device by modeling an equivalent electrical circuit. It has been already used to study RS mechanisms in NiO, $TiO_2$, $HfO_2$ and amorphous Si based structures [30,31]. In particular, it was used as a tool to characterize the nature of conductive nanofilaments, as it permits to discriminate between "complete" filaments bridging both electrodes (ohmic resistors) and "incomplete" filaments, where exists a gap between the tip of the filament and the nearby electrode. In the latter case, the electrical behavior of the oxide in the gap is modeled by a capacitor in parallel with a resistor [30,32]. Recently, impedance spectroscopy was used in memristive $SiO_x$-based systems to disclose the involved transport and switching mechanisms [33].

Figure 2 displays the complex impedance spectra Z'' vs. Z' (where Z'' and Z' are the imaginary and real components of the complex impedance **Z**, respectively) associated to the $R_{LOW}$, $R_{INT}$ and $R_{HIGH}$ states. In the three cases we observed semicircle-like curves, which can be accurately fitted by assuming an equivalent circuit constituted by a resistor ($R_1$) in series with two parallel resistor and capacitor combinations ($R_2//C_2$ and $R_3//C_3$), as shown in the sketch of Figure 3(a). We stress that no good fitting was obtained for other simpler equivalent circuits such as a single parallel resistor/capacitor combination (even in series with another resistor). The series resistor $R_1$ is related to the shift to the right of the diagram with respect to the origin of Z' axis. The fitted parameters for each resistance level are shown in Table 1. As depicted in Figure 3(b), the obtained equivalent circuit is consistent with the existence of a metallic filament which does not connect both electrodes [30] and produced as a consequence of the migration of $Ti^+$ due to the positive polarity of the forming protocol. This Ti filament is represented by the series resistor $R_1$, while the parallel resistor/capacitor elements correspond to the manganite layer between the tip of the Ti filament and the native $SiO_x$, and to the $SiO_x$ layer respectively. The metallic filament is consistent with the lack of electrode area dependence of $R_{LOW}$ state [28], while the existence of the gap between the tip of the filament and the bottom electrode is supported by the semiconductor-like temperature dependence of $R_{LOW}$ (see Ref. [28] and Supplementary Information).

|  | $R_1$ (Ω) | $R_2$ (Ω) | $C_2$ (nF) | $R_3$ (Ω) | $C_3$ (nF) |
|---|---|---|---|---|---|
| $R_{HIGH}$ | 800 | 1.5k | 4.5 | 85k | 0.7 |
| $R_{LOW}$ | 400 | 1.5k | 4.5 | 2.3k | 0.9 |
| $R_{INT}$ | 600 | 2k | 3.5 | 2.3k | 2 |

Table 1: Fitted circuit parameters for the three resistance levels.

The observation of Table 1 shows that the three resistance states are nicely reflected in the fitted resistances values: the $R_{HIGH}$ state is dominated by $R_3$ ~85kΩ (which we attribute to the $SiO_x$ layer after the positive polarity forming process). This value significantly drops to $R_3$~1.5kΩ in the transition to $R_{LOW}$. Due to the positive polarity of the stimulus the $SiO_x$ layer becomes oxygen deficient giving rise to a lower value of $R_3$ (see below for more details). The transition from $R_{LOW}$ to $R_{INT}$ is dominated by the variation of $R_2$, which changes from ~1.5kΩ to ~2kΩ, indicating a change of the resistance of the manganite layer. There are concomitant changes of the capacitances $C_1$ from 4.5 to 3.5 nF and $C_2$ from 0.9 nF to 2nF. Changes in the capacitance after the application of electrical stress has been observed in $LaAlO_3$-based memristive systems and were related to modifications of the interfacial depletion layers due to oxygen movement [34]. Further studies are needed in our case to confirm this possible scenario. The slight variations in $R_1$ could be related to some minor filament reshaping produced during the resistive transitions.

More information regarding the physics of the observed resistive transitions can be achieved by analyzing the dynamic I-V curves corresponding to the three resistive states, shown in Figure 4(a) and recorded by applying stimuli (current again) in the "read" range, that is, not high enough to trigger the resistive transitions. An easy way to graphically analyze the I-V curves consists of plotting the power exponent γ as a function of $V^{1/2}$, where γ is defined as $d(ln(I))/dln(V)$, as shown in Figures 4(b)-(d). This is better than performing fits of the I-V curves with the usual non-linear mathematical expressions related to each particular conduction mechanism through metal-oxide interfaces, as a constant value or a straight line dependence of γ with $V^{1/2}$ may reveal the conduction mechanism, without the need of judging numerically the quality of a fit. [35]. This is particularly the case for both $R_{HIGH}$ and $R_{LOW}$ were a constant value γ~3/2 is observed, indicating a dominating Child-Langmuir (Ch-L) conduction mechanism [36]. A similar mechanism has been observed in the case of $Al/SiO_2/n$-Si devices. [37]. In the Ch-L conducting regime, the current can be expressed as:

$$I = \frac{AV^{3/2}}{d^2} \quad (1)$$

where *A* is related to the permittivity of vacuum and with the electron's mass and charge, and d is the distance of the voltage drop. The switching from $R_{HIGH}$ to $R_{LOW}$ can then be ascribed to a decrease of *d* to ~*d*/10, indicating the formation of a conducting channel that dramatically reduces the effective value of *d*, but preserves the dominant conduction mechanism.

The graphic power exponent method is especially useful for the cases where two or more conduction mechanisms are present [38]. This is particularly what happens for the $R_{INT}$ state, which shows a non-monotonic evolution of γ, displaying a characteristic shape that includes an almost constant initial value of 1 at low voltages, a linear increase and a cusp with γ~4 for $V^{1/2}$~2.2 $V^{1/2}$. As shown in Ref. [38], this shape of the γ curve indicates the existence of a PF emission, which dominates the conduction mechanism in the intermediate current-voltage regime, in parallel with an ohmic element, and both in series with a lower γ process, visible at higher voltages, possibly related to the Ch-L conduction observed for $R_{HIGH}$ and $R_{LOW}$.

We recall that the PF conduction mechanism is a bulk process related to the electronic emission of carriers from traps in the oxide [36]. It can be found typically in interfaces of metal-complex oxides, like cuprates or cobaltites [39,40]. In this way, its origin can be related to the LCMO layer, which contributes with a pure ohmic conduction in the $R_{HIGH}$ and $R_{LOW}$ regimes, but becomes Poole-Frenkel after the $R_{LOW}$→$R_{INT}$ transition, due to the generation of defects (traps).

Within this framework, we propose the following physical scenario, as shown in Figure 3(b): after the initial positive electroforming an incomplete metallic filament is formed and the device is stabilized in the $R_{HIGH}$ resistance level, where the device resistance is dominated by the $SiO_x$ layer (with *x* close to the stochiometric value 2) and the Ch-L conduction mechanism originating in this layer prevails. The application of positive stimulus leads into the SET transition ($R_{HIGH}$ to $R_{LOW}$), which is related to oxygen transfer from the $SiO_x$ to the manganite layers. In the case of $SiO_x$, it is known that the electrical conductance is enhanced if the oxygen stochiometry is reduced. This is reflected by the abrupt decrease in $R_3$. The conduction mechanism in the $R_{LOW}$ state is still dominated by the $SiO_x$ layer, remaining Ch-L type but with an enhanced conducting channel, consequence of its lower oxygen stochiometry. Upon the application of further positive stimulus, more oxygen ions are transferred from the $SiO_x$ to the manganite layer, stabilizing the $R_{INT}$ state. In this state, the conduction mechanism starts being dominated by the manganite layer. It can be argued that the transfer of oxygen ions from the $SiO_x$ to the manganite layer fill oxygen vacancies and above a certain threshold ($R_{LOW}$→$R_{INT}$, reflected by an increase in $R_2$) forms trap-centers, increasing the manganite resistance and stabilizing the Poole-Frenkel conduction mechanism that dominates the macroscopic transport. The application of negative stimulus reverses the processes already described; oxygen is transferred back from the manganite to the SiO$_x$ layer which becomes more stochiometric, returns to R$_{HIGH}$ and the dominating conduction mechanism become Ch-L again.

In summary, we have conclusively disclosed the physical scenario associated to three-level resistive switching in n-Si/SiO$_x$/LCMO/Ti devices. The contribution of three regions (metallic filament, manganite and SiO$_x$ layers) to the transport properties has been individualized. Our results show that the existence of the native SiO$_x$ layer plays a key role in the stabilization of an intermediate resistance level, indicating that the combination of two or more active RS oxides in single devices improves their functionalities. We understand that these multilevel devices are interesting and promising as their fabrication procedure is rather simple and they are fully compatible with standard Si-based electronics.

Supplementary Material

See supplementary material for the X-ray photoemission spectroscopy characterization of an as received Si substrate and the temperature dependence of R$_{LOW}$, R$_{HIGH}$ and R$_{INT}$.

We acknowledge financial support from CONICET (PIP 291), PICT "MeMO"(0788) and CIC-Buenos Aires. We thank U. Lüders and J. Lecourt, from CRISMAT, for the preparation of the manganite target, and P. Granell, from INTI, for the SEM-FIB measurements. We also thank S. Bengió, from CAB-Bariloche, for the XPS measurement.

Figure Captions

Figure 1: (a) Voltage-current curve corresponding to a Ti/LCMO/SiO$_2$/n-Si device; (b) Hysteresis Switching Loops corresponding to the same device. Three non-volatile resistance levels are seen.

Figure 2: (a),(b): Complex impedance spectra corresponding to the three resistance levels. Full red lines correspond to the fittings of the spectra.

Figure 3: (a) Equivalent circuit proposed for modeling the complex impedance spectra; (b) Sketch showing the physical scenario related to the equivalent circuit.

Figure 4: (a) Current-voltage curves corresponding to the three resistance levels for low stimulus ("reading" range); (b),(c),(d) Evolution of $\gamma = d(\ln(I))/d\ln(V))$ as a function of $V^{1/2}$, for the three resistance levels.

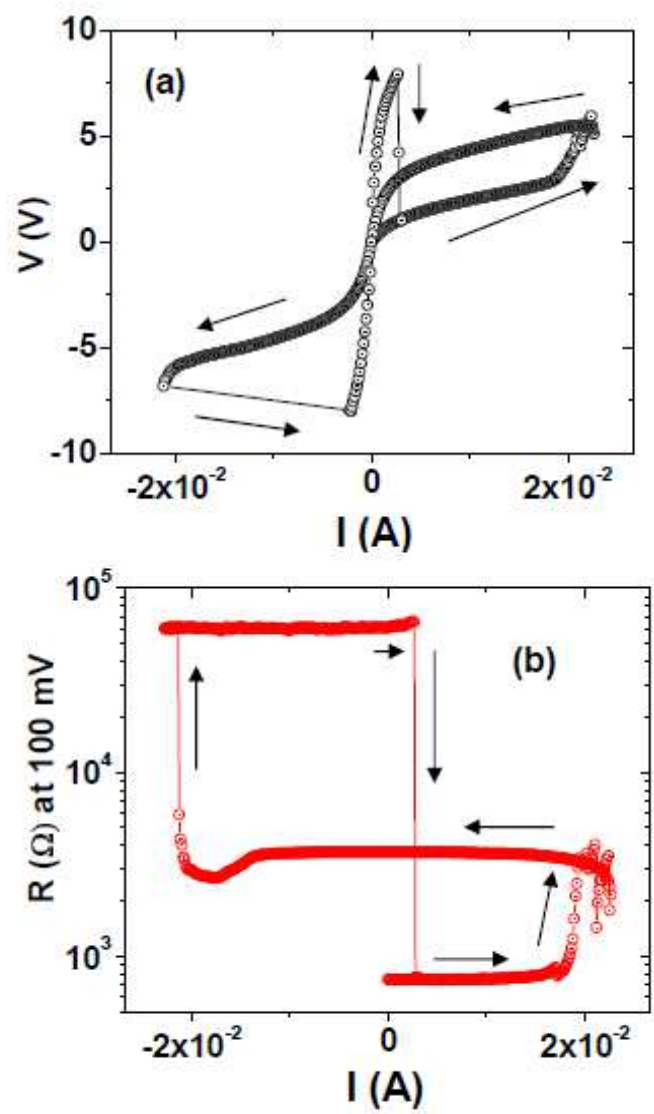

**Figure 1**

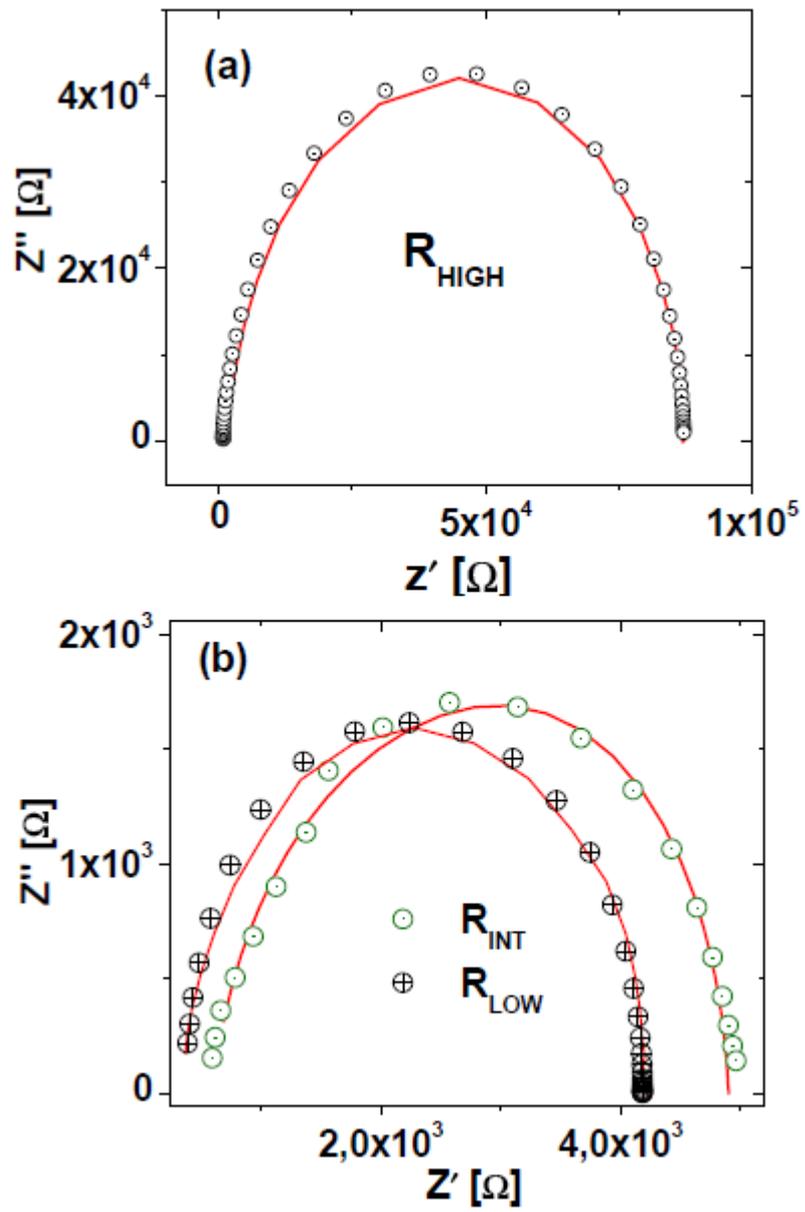

**Figure 2**

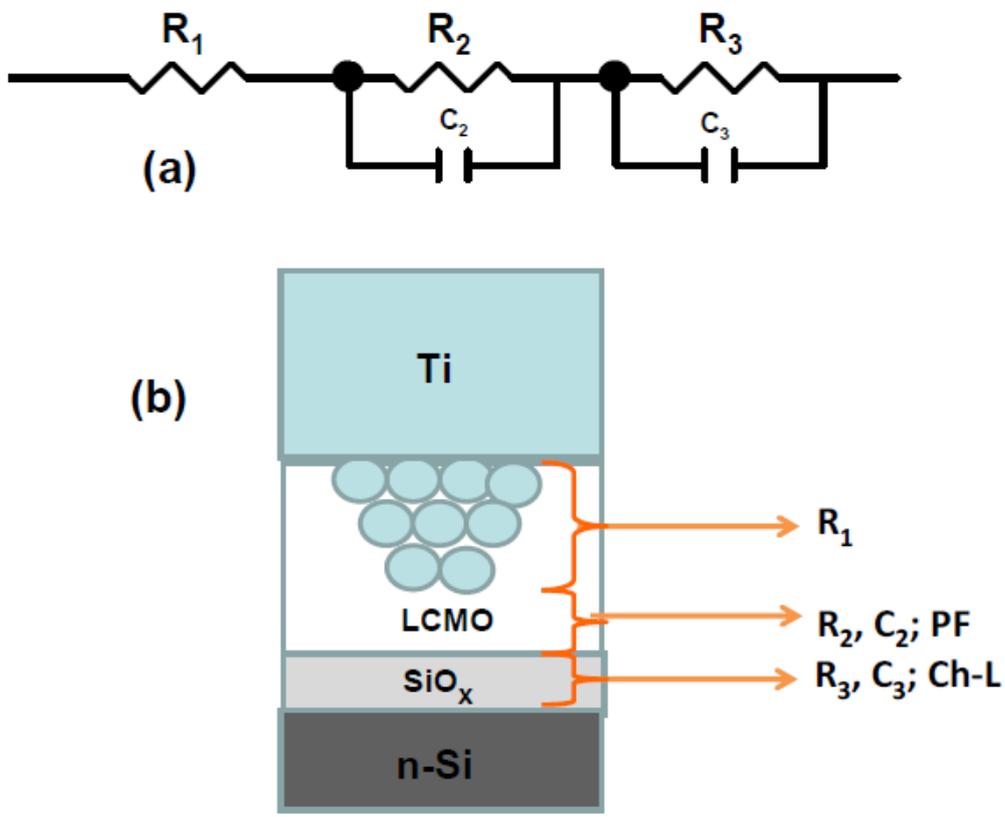

**Figure 3**

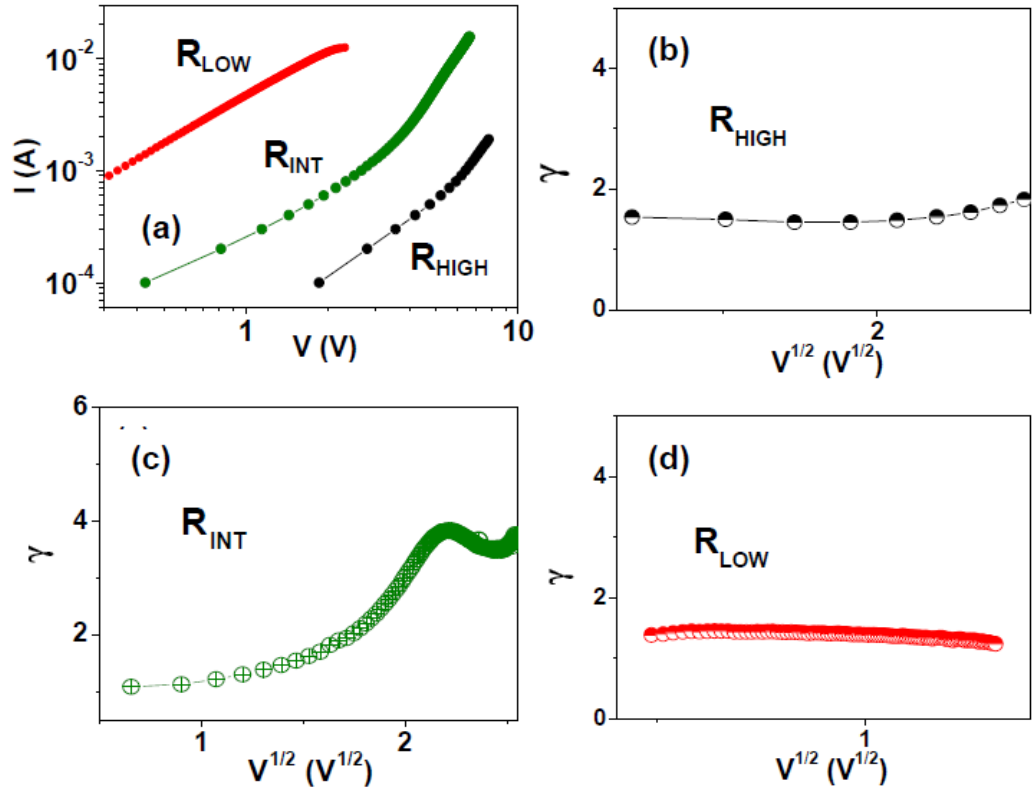

**Figure 4**